\documentclass[12pt] {article}
\usepackage{amstex}

\def\agt{
\mathrel{\raise.3ex\hbox{$>$}\mkern-14mu\lower0.6ex\hbox{$\sim$}}
}
\def\alt{
\mathrel{\raise.3ex\hbox{$<$}\mkern-14mu\lower0.6ex\hbox{$\sim$}}
}

\begin {document}

\bibliographystyle{unsrt}    

\vspace{3mm} 
\Large
{\centerline{\bf Fermion-boson duality in integrable}}
\vspace{3mm} 
{\centerline{\bf quantum field theory}
\vspace{2mm}

\large

\vspace{7mm}

\centerline {\bf P.~Baseilhac, V. A. Fateev\,\footnote{On leave of absence from L. D. Landau Institute for Theoretical Physics, ul. Kosygina 2, 117940 Moscow, Russia}}
\small\normalsize

\vspace{3mm}

\centerline {Laboratoire de Physique Math\'ematique, Universit\'e Montpellier II}
\centerline {Place E.~Bataillon, 34095 Montpellier, France}
\vspace{3mm}

\vspace{6mm}
\large
\begin{abstract}
We introduce and study one parameter family of integrable quantum field theories. This family has a Lagrangian description in terms of massive Thirring fermions $\psi,\psi^{\dagger}$ and charged bosons $\chi,\overline{\chi}$ of complex sinh-Gordon model coupled with $BC_n$ affine Toda theory. Perturbative calculations, analysis of the factorized scattering theory and the Bethe ansatz technique are applied to show that under duality transformation, which relates weak and strong coupling regimes of the theory the fermions $\psi,\psi^{\dagger}$ transform to bosons  and $\chi,\overline{\chi}$ and vive versa. The scattering amplitudes of neutral particles in this theory coincide exactly with S-matrix of particles in pure $BC_n$ Toda theory, i.e. the contribution of charged bosons and fermions to these amplitudes exactly cancel each other. We describe and discuss the symmetry responsible for this compensation property.
\end{abstract}
\normalsize
\newpage
\section{Introduction}\label{introduction} 

\ \ \ \ \ Duality plays an important role in the analysis of statistical and quantum field theory (QFT) systems. It maps a weak coupling region of one theory to a strong coupling region of the other and makes it possible to use perturbative and semiclassical methods for the study of dual systems in different regions of the coupling constants. For example, the well-known duality between the sine-Gordon and massive Thirring models \cite{1} plays a crucial role for study of many two-dimensional quantum systems. The phenomenon of electric-magnetic duality in four-dimensional gauge theories, conjectured in \cite{2} and developped in \cite{3} opens the possibility for the non-perturbative analysis of the spectrum and the phase structure in supersymmetric Yang-Mills theory.

Known for many years the phenomenon of duality in QFT still looks rather mysterious and needs further study. This analysis essentially simplifies for two-dimensional integrable relativistic theories. These QFT's besides the Lagrangian formulation possess also an unambiguous definition in terms of factorized scattering theory (FST). The FST, i.e. the explicit description of the spectrum of particles and their scattering amplitudes, contains all information about QFT. These data permit one to use non-perturbative methods for the analysis of integrable QFT and makes it possible in some cases to justify the existence of two different (dual) representations for the Lagrangian description of the theory. An interesting example of duality in two-dimensional integrable systems is the weak coupling - strong coupling flow from the affine Toda theories (ATT) to the same theories with the dual affine Lie algebra \cite{4}. The duality in rank $r$ non-simply laced ATTs coupled with massive Thirring model was studied in \cite{7}. It was shown there that dual theory can be formulated as the non-linear sigma-model with Witten's Euclidian black hole metric \cite{6} (complex sinh-Gordon theory) coupled with non-simply laced ATTs. Lie algebras of these ``dual'' ATTs belong to the dual series of affine algebras but have the smaller rank $\tilde{r}=r-1$.

In this paper in section 2 we consider one-parameter family of integrable QFT, which has the Lagrangian formulation in terms of complex fermion field ($\psi, \psi^{\dagger}$), complex boson field ($\chi,\overline{\chi}$) and $n$ scalar fields $\varphi=(\varphi_1,...,\varphi_n)$. This QFT possesses $U(1) \otimes U(1)$ symmetry generated by fermion and boson charges $Q_{\psi}$ and $Q_{\chi}$. It can be considered as $BC_n$ ATT coupled with massive Thirring and complex sinh-Gordon (CSG) \cite{5} models. In the weak coupling region this QFT admits a perturbative analysis. There the spectrum of particles, besides the charged fermions ($\psi, \psi^{\dagger}$) and bosons ($\chi,\overline{\chi}$), contains the scalar neutral particles $M_a$ with the masses characteristic for the $BC_n$ ATT. Perturbative calculations show that classical mass ratios are not destroyed by quantum corrections and that charged particles possess non-diagonal scattering. The scattering amplitudes of the charged particles can be expressed through the solution of factorization (Yang-Baxter) equation. The perturbative analysis together with $U(1) \otimes U(1)$ symmetry fix this solution up to one parameter, which depends on the coupling constant. 

In section 3 we introduce the external field $A$ coupled with charges $Q_{\psi}$ and $Q_{\chi}$. We use standard Bethe ansatz (BA) technique to find the exact relation between the coupling constant and parameter of FST. We show that in the strong coupling regime the behavior of fermions $\psi$ in the external field is similar to the behavior of weakly coupled bosons $\chi$ and vice versa. The resulting FST possesses the property of self-duality together with fermion-boson transformation $\psi \leftrightarrow \chi$. Another remarkable property of this QFT is the exact coincidence of the scattering amplitudes $S_{ab}$ of neutral particles $M_a$ with S-matrix of pure $BC_n$ ATT. It means that fermion and boson contributions to these amplitudes exactly cancel each other. In the last section we describe and discuss the symmetry which is responsible for this compensation property.

\section{Integrable deformation of $BC_n$ Toda theory and factorized scattering theory}

\ \ \ \ \ In this section we consider the QFT which can be described by Dirac fermion $\psi$, complex scalar field $\chi$ and $n$ scalar fields $\varphi=(\varphi_1,...,\varphi_n)$ with the action :
\begin{eqnarray} 
{\cal A}_n &=&  \int d^2x\ \Big[ \frac{1}{2}\frac{\partial_{\mu}\overline{\chi}\partial_{\mu}\chi}{1+(\frac{\beta}{2})^2|\chi|^2} + i\overline{\psi} \gamma_{\mu}\partial_{\mu}\psi -\frac{g}{2}(\overline{\psi}\gamma_{\mu}\psi)^2 - \frac{M_0^2}{2}|\chi|^2e^{\beta\varphi_1} - M_0\overline{\psi} \psi e^{-\beta\varphi_n}\nonumber \\
&&+\frac{1}{2}(\partial_{\mu}\varphi)^2 - \frac{M_0^2}{2\beta^2}\big[ 2\exp(\beta \varphi_1) + 2\sum_{i=1}^{n-1}\exp\beta(\varphi_{i+1}-\varphi_{i}) + \exp(-2\beta\varphi_n) \big] \Big],\label{1}
\end{eqnarray}
where\ \ \ \ $g/\pi = -\frac{\beta^2}{4\pi(1 + \beta^2/4\pi)}$.

The last term $-\frac{M_0^2}{2\beta^2}\exp(-2\beta\varphi_n)$  in (\ref{1}) plays the role of usual contact counterterm which cancels divergencies coming from fermion loop. With this term the action (\ref{1}) has the form of $BC_n$ affine Toda theory coupled with massive Thirring and CSG \cite{5} models. Following the notations of ref. \cite{7} we denote this QFT as ${\cal{BC}}_n(\psi,\chi,\beta)$. It possesses $U(1) \otimes U(1)$ symmetry, generated by the charges 
\begin{eqnarray} 
Q_\psi = \int dx \overline{\psi}\gamma_{0}\psi\ \ \ ;\ \ \ 
 \ \ \  
Q_\chi = \int dx \frac{\overline{\chi}\partial_0\chi-\chi\partial_0\overline{\chi}}{2i(1+(\frac{\beta}{2})^2|\chi|^2)}.\label{2} 
\end{eqnarray}

The QFT (\ref{1}) is integrable. It possesses the local integrals $P_s$ with odd (Lorentz) integer spins $s$. The explicit form of these integrals is not in the scope of this paper. Some additional symmetry of ${\cal{BC}}_n(\psi,\chi,\beta)$ model generated conserved charges with half-integer spin $s=n+\frac{1}{2}$ is described in the last section. We also checked at the tree level that multiparticles amplitudes factorize into two-particle ones. 

For small $\beta$ we can use the perturbation theory for the analysis of QFT (\ref{1}). Its spectrum contains charged fermions ($\psi,\psi^{\dagger}$), charged  bosons ($\chi,\overline{\chi}$) with mass $M$ and neutral particles $M_a,\  a=1,...,n$. In one loop approximation the mass ratios in ${\cal{BC}}_n(\psi,\chi,\beta)$ theories are not destroyed by the quantum corrections and have the classical values :
\begin{eqnarray}
M_a=2M\sin(\frac{\pi a}{h});\ \ \ \ a=1,...,n;\label{3}
\end{eqnarray}
here and later $h=2n+1$.

The scattering theory in integrable theory (\ref{1}) is completly defined by a two particle S-matrix. The non-diagonal scattering is possible only between the particles $\psi,\psi^{\dagger},\chi,\overline{\chi}$ of equal mass. All other amplitudes $S_{a\chi}$, $S_{a\psi}$ and $S_{ab}$ are the pure phases. The scattering matrix of charged particles with $U(1) \otimes U(1)$ symmetry ($\psi \rightarrow e^{i\eta}\psi$, $\chi \rightarrow e^{i\xi}\chi$) and $\mathbb{C}, \mathbb{P}, \mathbb{T}$ invariance is characterized by the following amplitudes :
\begin{eqnarray} 
&&|\psi(\theta_1)\psi(\theta_2)>_{in}=S_\psi(\theta)|\psi(\theta_2)\psi(\theta_1)>_{out}\ ; \ \ |\chi(\theta_1)\chi(\theta_2)>_{in}=S_{\chi}(\theta)|\chi(\theta_2)\chi(\theta_1)>_{out}\ ; \nonumber\\
&&|\psi(\theta_1)\chi(\theta_2)>_{in}=\gamma(\theta)|\chi(\theta_2)\psi(\theta_1)>_{out} +\  \delta(\theta)|\psi(\theta_2)\chi(\theta_1)>_{out}\ ;\nonumber \\
&&|\psi^{\dagger}(\theta_1)\chi(\theta_2)>_{in}=\alpha(\theta)|\chi(\theta_2)\psi^{\dagger}(\theta_1)>_{out} +\  \beta(\theta)|\psi^{\dagger}(\theta_2)\chi(\theta_1)>_{out}\ ;\nonumber \\
&&|\psi(\theta_1)\psi^{\dagger}(\theta_2)>_{in}= T_\psi(\theta)|\psi^{\dagger}(\theta_2)\psi(\theta_1)>_{out} +\ R_\psi(\theta)|\psi(\theta_2)\psi^{\dagger}(\theta_1)>_{out}\label{4} \\ 
 &&\ \ \ \ \ \  \ \ \ \ \ \ \ \ \ \ \ \ \ \ \ \ \ \ \ \ \ \ \ \ \ \ +\ \mu(\theta)|\chi^{\dagger}(\theta_2)\chi(\theta_1)>_{out} + \nu(\theta)|\chi(\theta_2)\chi^{\dagger}(\theta_1)>_{out}\ ;\nonumber\\
&&|\chi(\theta_1)\chi^{\dagger}(\theta_2)>_{in}= T_\chi(\theta)|\chi^{\dagger}(\theta_2)\chi(\theta_1)>_{out} +\ R_\chi(\theta)|\chi(\theta_2)\chi^{\dagger}(\theta_1)>_{out} \nonumber\\
 && \ \ \ \ \ \  \ \ \ \ \ \ \ \ \ \ \ \ \ \ \ \ \ \ \ \ \ \ \ \ \ \ +\ \mu(\theta)|\psi^{\dagger}(\theta_2)\psi(\theta_1)>_{out} +\ \nu(\theta)|\psi(\theta_2)\psi^{\dagger}(\theta_1)>_{out}\nonumber.
\end{eqnarray}
All the amplitudes (\ref{4}) depend on rapidity difference $\theta=\theta_1-\theta_2$. They satisfy the following crossing symmetry condition :
\begin{eqnarray}
&& S_{\psi}(i\pi-\theta)=T_\psi(\theta);\ \ S_{\chi}(i\pi-\theta)=T_\chi(\theta);\ \ R_{\psi}(i\pi-\theta)=R_\psi(\theta); \label{5} \\
&& R_{\chi}(i\pi-\theta)=R_\chi(\theta);\ \ \alpha(i\pi-\theta)=\gamma(\theta);\ \ \beta(i\pi-\theta)=\mu(\theta);\ \ \delta(i\pi-\theta)=\nu(\theta).\nonumber
\end{eqnarray}

The perturbative expansion for these amplitudes in the first order in $\beta^2$ has the form:
\begin{eqnarray}
&& -S_{\psi}(\theta) = 1 + \frac{i\beta^2}{4h} \Big[ h\coth\big(\frac{h\theta}{2}\big)-\frac{2}{\sinh\theta}\Big]  + {O}(\beta^4) = -T_\psi(i\pi-\theta)\ ;\nonumber \\
&& S_{\chi}(\theta) = 1 - \frac{i\beta^2}{4h} \Big[ h\coth\big(\frac{h\theta}{2}\big)+\frac{2}{\sinh\theta}\Big]  + {O}(\beta^4) = T_\chi(i\pi-\theta)\ ;\nonumber \\
&& R_\psi(\theta) = R_\chi(\theta) =  - \frac{i\beta^2}{2\sinh(h\theta)} + {O}(\beta^4)\ ;\nonumber \\
&& \alpha(\theta) = \gamma(\theta) = 1 - \frac{i\beta^2}{2h\sinh(\theta)} + {O}(\beta^4)\ ;\label{6} \\ 
&& \beta(\theta) = \delta(\theta) = \mu(i\pi-\theta) = \nu(i\pi-\theta) = - \frac{i\beta^2}{4\sinh(h\theta/2)} + {O}(\beta^4).\nonumber 
\end{eqnarray}

Factorization property imposes non-trivial limitations to the scattering amplitudes. They should satisfy the functional Yang-Baxter (factorization) relations. There are two types of  $\mathbb{C}, \mathbb{P}, \mathbb{T}$ invariant solutions of Yang-Baxter equations with $U(1) \otimes U(1)$ symmetry. The first one corresponds to the case $S_\psi=S_\chi$ and is expressed through the direct product of two sine-Gordon S-matrix. It follows from eq. (\ref{6}) that in the ${\cal{BC}}_n(\psi,\chi,\beta)$ model is realized another case $S_\psi\neq S_\chi$. The solution of the factorization equation with this property possesses one (besides scale of $\theta$) arbitrary parameter $x$ and has the form :
\begin{eqnarray}
&& S_\psi(\theta)= - \frac{\sinh(\lambda \theta +ix\pi)}{\sinh(\lambda \theta)}Y_n(\theta); \ \ \ S_\chi(\theta)=  \frac{\sinh(\lambda \theta -ix\pi)}{\sinh(\lambda \theta)}Y_n(\theta); \nonumber \\
&& T_\psi(\theta)= - \frac{\cosh(\lambda \theta -ix\pi)}{\cosh(\lambda \theta)}Y_n(\theta); \ \ \ T_\chi(\theta)=  \frac{\cosh(\lambda \theta +ix\pi)}{\cosh(\lambda \theta)}Y_n(\theta);\label{7} \\
&& R_\psi(\theta)=R_\chi(\theta)= -\frac{2i\sin x\pi}{\sinh(2\lambda\theta)}Y_n(\theta); \ \ \ \alpha(\theta)=\gamma(\theta)=Y_n(\theta); \nonumber \\
&& \beta(\theta)=\delta(\theta)=-i\frac{\sin x\pi}{\sinh(\lambda\theta)}Y_n(\theta); \ \ \ \mu(\theta)=\nu(\theta)= p\frac{\sin x\pi}{\cosh(\lambda\theta)}Y_n(\theta),\ \ p^2=1. \nonumber
\end{eqnarray}
The solution (\ref{7}) is consistent with crossing symmetry (\ref{5}) only if 
\begin{eqnarray}
\lambda=\frac{h}{2}=n+\frac{1}{2}, \ \ \ p=(-1)^{n+1}.\label{8}
\end{eqnarray}
Function $Y_n(\theta)$ satisfies the unitarity and crossing symmetry relations :
\begin{eqnarray}
Y_n(\theta)Y_n(-\theta) = \frac{\sin^2\pi x}{\sinh^2\theta+\sin^2\pi x}\ ;\ \ Y_n(\theta)=Y_n(i\pi-\theta).\label{9}
\end{eqnarray}
The minimal solution of eqs. (\ref{9}) is known \cite{8} and has the form :
\begin{eqnarray}
Y_n(\theta) &=& R(\theta)R(i\pi-\theta)\ ; \ \ R(\theta)=\prod_{l=0}^{\infty}\frac{F_l(\theta,x)F_l(\theta,1-x)}{F_l(\theta,0)F_l(\theta,1)}; \nonumber \\
&&F_l(\theta,x) = \frac{\Gamma(\frac{h\theta}{4\pi i}+\frac{hl}{2}+x)}{\Gamma(\frac{h\theta}{4\pi i}+\frac{h}{2}(l+\frac{1}{2})+x)}.\label{10}
\end{eqnarray}
In particular the amplitudes $S_\psi$ and $S_\chi$ which are the pure phases can be represented as :
\begin{eqnarray}
-S_\psi(\theta)&=& \exp \Big[ i\int_{-\infty}^{\infty} \frac{d\omega}{\omega} \frac{\sin(\omega\theta) \sinh(\pi x\omega/h) \cosh[\pi\omega(h+2(1-x))/2h]}{\sin(\pi\omega/h)\cosh(\pi\omega/2)}\Big]; \label{11}\\
-S_\chi(\theta)&=& \exp \Big[ i\int_{-\infty}^{\infty} \frac{d\omega}{\omega} \frac{\sin(\omega\theta) \sinh(\pi(1-x)\omega/2h) \cosh[\pi\omega(h+2x)/2h]}{\sin(\pi\omega/h)\cosh(\pi\omega/2)}\Big].\nonumber
\end{eqnarray}
We note that function $Y_n(\theta)$ is invariant under the transformation $x \rightarrow 1-x$. S-matrix (\ref{7}) is invariant under this transformation together with fermion-boson transformation $\psi \leftrightarrow \chi$.

For small $x$ function $Y_n(\theta)$ has the expansion :
\begin{eqnarray}
Y_n(\theta)=1-\frac{2\pi i x}{h\sinh\theta} + {O}(x^2). \label{12}
\end{eqnarray}
We can see that small $x$ expansion of amplitudes (\ref{7}) coincides with perturbative expansion (\ref{6}) if 
\begin{eqnarray}
x=\frac{\beta^2}{4\pi}\big(1 + {O}(\beta^2)\big).\label{13}
\end{eqnarray}

The amplitudes (\ref{7}) possess poles at the physical strip $0<Im\theta<\pi$ which are located at the points $\theta_a$, where :
\begin{eqnarray}
\theta_a = i\pi(1-\frac{2a}{h}), \ \ \  a=1,...,n. \label{14}
\end{eqnarray}
These poles correspond to the neutral bound states $M_a$ with masses (\ref{3}). The scattering amplitudes $S_{a\psi}$, $S_{a\chi}$, $S_{ab}$ including these particles can be obtained by the usual fusion procedure. For this purpose it is convenient to represent the particles $M_a$ in the form :
\begin{eqnarray}
M_a(\theta) &=&\psi(\theta+\frac{\theta_a}{2})\psi^{\dagger}(\theta-\frac{\theta_a}{2})+\psi^{\dagger}(\theta+\frac{\theta_a}{2})\psi(\theta-\frac{\theta_a}{2}) \nonumber \\
&+&(-1)^a \Big[ \chi(\theta+\frac{\theta_a}{2})\chi^{\dagger}(\theta-\frac{\theta_a}{2})+\chi^{\dagger}(\theta+\frac{\theta_a}{2})\chi(\theta-\frac{\theta_a}{2})\Big],\label{15}
\end{eqnarray}
and use for the particles $\psi(\theta)$, $\chi(\theta)$ the commutation relations with S-matrix (\ref{7}). In this way we obtain the following expression for the amplitudes $S_{a\psi}$, $S_{a\chi}$ :
\begin{eqnarray}
S_{a\psi}(\theta)=S_{a\chi}(\theta)=S_{\psi}(\theta+\frac{\theta_a}{2})T_{\psi}(\theta-\frac{\theta_a}{2}) &+& R_{\psi}(\theta+\frac{\theta_a}{2})R_{\psi}(\theta-\frac{\theta_a}{2})\label{16}\\
 &&\ \ \ \ \ \ \ + \  2\beta(\theta+\frac{\theta_a}{2})\mu(\theta-\frac{\theta_a}{2}).\nonumber 
\end{eqnarray}
In particular the amplitudes $S_{1\psi}=S_{1\chi}$ can be written as :
\begin{eqnarray}
S_{1\psi}(\theta)=S_{1\chi}(\theta)=\frac{\sinh\theta - i\cos(\pi(1-2x)/h)}{\sinh\theta + i\cos(\pi(1-2x)/h)}\frac{\sinh\theta + i\cos(\pi/h)}{\sinh\theta - i\cos(\pi/h)}.\label{17}
\end{eqnarray}
To describe all two particle amplitudes we introduce the notations \cite{4} :
\begin{eqnarray}
(z) &=& \frac{\sinh(\frac{\theta}{2}+\frac{i\pi z}{2h})}{\sinh(\frac{\theta}{2}-\frac{i\pi z}{2h})}\ \  , \ \ \ \lbrace z \rbrace = \frac{(z-1)(z+1)}{(z-1+2x)(z+1-2x)},\label{18}
\end{eqnarray}
then 
\begin{eqnarray}
S_{a\psi}=S_{a\chi}(\theta)=\prod_{p=1}^a \lbrace \frac{h}{2}+2p-a-1\rbrace.\label{19}
\end{eqnarray}
The scattering amplitudes of neutral particles $M_a$ have the form :
\begin{eqnarray}
S_{ab}(\theta) = \prod^{|a+b|-1}_{|a-b|+1 (step 2)}\lbrace p\rbrace\lbrace h-p\rbrace.\label{20}
\end{eqnarray}
The amplitudes (\ref{19}), (\ref{20}) are invariant under the transformation $x \rightarrow 1-x$. They also are in agreement with first order of perturbation theory if $x$ has the form (\ref{13}). To find the exact relation between parameter $x$ and coupling constant $\beta$ in action (\ref{1}) we have to use non-perturbative approach to the QFT ${\cal{BC}}_n(\psi,\chi,\beta)$.

\section{Non-perturbative consideration}
\ \ \ \ The QFT (\ref{1}) possesses the symmetry $U(1)\otimes U(1)$ generated by the charges $Q_\psi$ and $Q_\chi$ (\ref{2}) and admits the introduction of the external fields $A_\psi$ and $A_\chi$ coupled with these charges. For simplicity we consider the configurations with only one non-zero field $A_\psi$ or $A_\chi$, which we denote as $A$. The hamiltonian ${\cal{H}}_\psi ({\cal{H}}_\chi)$ in the external field $A$ has an additional term equal to $-AQ_\psi (-AQ_\chi )$ : \begin{eqnarray}
{\cal{H}}_\psi = {\cal{H}}_{0} - AQ_\psi\ \ ;\ \ {\cal{H}}_\chi = {\cal{H}}_{0} - AQ_\chi, \label{21}
\end{eqnarray}
where ${\cal{H}}_{0}$ is the Hamiltonian of the QFT (\ref{1}).

To find the exact relation between $x$ and $\beta$ we calculate the specific ground state energy ${\cal{E}}_\psi(A) ({\cal{E}}_\chi(A))$ in the limit $A \rightarrow \infty$ from the Hamiltonian and from S-matrix data. The calculation of these asymptotics from the Hamiltonian (\ref{21}) follows exactly the lines of ref. \cite{7}, where similar calculations were done, so we reproduce here the result :
\begin{eqnarray}
{\cal{E}}_\psi(\beta, A \rightarrow \infty) = -\frac{A^2(1+\beta^2/4\pi)}{2\pi},\label{22}\\
{\cal{E}}_\chi(\beta, A \rightarrow \infty) = -\frac{2A^2(1+\beta^2/4\pi)}{\beta^2}.\nonumber
\end{eqnarray}

We calculate now the same values from the S-matrix using the BA approach (see for example refs. \cite{9}). We consider the case corresponding to the ground state energy ${\cal{E}}_\psi$ taking into account that for function ${\cal{E}}_\chi$ all consideration differs only by the notations ($\psi \rightarrow \chi$). Due to additional term $-AQ_\psi$ every positively (negatively) charged particle $\psi(\theta)(\psi^{\dagger}(\theta))$ acquires the additionnal energy $A(-A)$. For $A>M$ the ground state containes a sea of positively charged particles $\psi(\theta)$, which fill all possible states inside some interval $-B<\theta<B$. The distribution $\epsilon_\psi(\theta)$ of particles within this interval is determinated by their scattering amplitude $S_\psi(\theta)$. The specific ground state energy can be expressed through the function $\epsilon_\psi(\theta)$ as :
\begin{eqnarray}
{\cal{E}}_\psi(A)-{\cal{E}}_\psi(0)= -\frac{M}{2\pi}\int_{-B}^{B}\cosh\theta\epsilon_\psi(\theta)d\theta,\label{23}
\end{eqnarray}
where non-negative function $\epsilon_\psi(\theta)$ satisfies, in the interval $-B< \theta < B$, the BA equation:
\begin{eqnarray}
\int_{-B}^{B}\tilde{K}_\psi(\theta-\theta')\epsilon_\psi(\theta')d\theta'= A - M\cosh\theta,\label{24}
\end{eqnarray}
where the kernel $\tilde{K}_\psi(\theta)$ in (\ref{24}) is related to the $\psi \psi$ scattering phase by :
\begin{eqnarray}
\tilde{K}_\psi(\theta)=\delta(\theta) -\frac{1}{2\pi i} \frac{d}{d\theta} \log S_\psi(\theta),\label{25}
\end{eqnarray}
and the parameter $B$ is determinated by the boundary conditions $\epsilon_\psi(\pm B)=0$. 

The Fourier transform $K_\psi(\omega)$ of the kernel (\ref{25}) can be obtained from eq. (\ref{11}) and has the form :
\begin{eqnarray}
{K}_\psi(\omega)= \frac{\sinh\big[\pi \omega (1-x)/h\big]\cosh\big[\pi\omega(h+2x)/h\big]}{\cosh(\pi \omega/2)\sinh(\pi \omega/h)}.\label{26}
\end{eqnarray}
For the function ${\cal{E}}_\chi(\beta,A)$ we obtain exactly the same equations with function $\epsilon_\chi(\theta)$ satisfying eq. (\ref{24}), where kernel $\tilde{K}_\chi(\theta)$ is related to the amplitude $S_\chi$ by the eq. (\ref{25}). The amplitudes $S_\psi$ and $S_\chi$ (\ref{11}) are connected by the transformation $x \rightarrow 1-x$. It means that:
\begin{eqnarray}
{K}_\chi(\omega)= \frac{\sinh\big[\pi \omega x/h\big]\cosh\big[\pi\omega(h+2(1-x))/h\big]}{\cosh(\pi \omega/2)\sinh(\pi \omega/h)}.\label{27}
\end{eqnarray}

The main term of the asymptotics of the function ${\cal{E}}_\psi$ $({\cal{E}}_\chi)$ at $A \rightarrow \infty$ can be expressed explicitly through the kernel ${K}_\psi(\omega)$ $({K}_\chi(\omega))$ by the relation \cite{10} :
\begin{eqnarray}
{\cal{E}}_{\psi}(\beta, A \rightarrow \infty) &=& -\frac{A^2}{2\pi K_\psi(0)} = -\frac{A^2}{2\pi (1-x)}\ ;\label{28} \\
{\cal{E}}_{\chi}(\beta, A \rightarrow \infty) &=& -\frac{A^2}{2\pi K_\chi(0)} = -\frac{A^2}{2\pi x}.\nonumber
\end{eqnarray}
Comparing eqs. (\ref{22}) and (\ref{27}) we find the exact value for parameter $x(\beta)$ :
\begin{eqnarray}
x(\beta)=\frac{\beta^2}{4\pi+\beta^2}=1-x\big(\frac{4\pi}{\beta}\big).\label{29}
\end{eqnarray}

The term ${\cal{E}}_0$ in eq. (23) is the bulk vacuum energy of QFT (\ref{1}). It can also be expressed through the kernel ${K}_\psi(\omega)({K}_\chi(\omega))$ by the relation :
\begin{eqnarray}
{\cal{E}}_0=-\frac{M^2}{8}\big[{K}_{\psi,\chi}(\omega)\cosh\frac{\pi\omega}{2}\mid_{\omega=i}\big]^{-1}= \frac{M^2\sinh \pi/h}{8\sin(\pi x/h)\sin(\pi(1-x)/h)}. \label{30}
\end{eqnarray}
The bulk vacuum energy is symmetric under the transformation  $\beta \rightarrow 4\pi/\beta$. It follows from the eqs. (\ref{26}), (\ref{27}) and (\ref{29}) that the transformation relates the function ${\cal{E}}_\psi$ and ${\cal{E}}_\chi$ :
\begin{eqnarray}
{\cal{E}}_{\psi}(\beta,A)={\cal{E}}_{\chi}(\frac{4\pi}{\beta},A).\label{31}
\end{eqnarray}
It means that strong coupling behavior ($\beta>>1$) of the fermions $\psi,\psi^{\dagger}$ in the external field $A$ coincides with weak coupling behavior ($\beta<<1$) of the bosons $\chi,\chi^{\dagger}$ and vice versa. In the weak coupling region the behavior of functions ${\cal{E}}_\psi$ and ${\cal{E}}_\chi$ is rather different. In this limit $K_\psi(\theta)=1+{O}(\beta^2)$ and function ${\cal{E}}_\psi(A)$ can be easily calculated. As function of $\beta$ it has the smooth behavior at $\beta<<1$ and can be written in parametric form as :
\begin{eqnarray}
&&{\cal{E}}_{\psi}(A)-{\cal{E}}(0)=-\frac{A^2}{2\pi}\Big(    1-2u+2u(1-u)\log\Big(\frac{u}{1-u}\Big)    \Big)+{O}(\beta^2),\nonumber \\
&&\ \ \ \ \ \ \ \  \Big(\frac{M}{2A}\Big)^2=u(1-u).\label{32}
\end{eqnarray}
The kernel $K_\chi(\theta)$ in the weak coupling limit is not trivial and has the form :
\begin{eqnarray}
K_\chi(\theta)\simeq \frac{\beta^2\omega}{4h}\frac{\cosh\big[\pi\omega(h+2)/h\big]}{\cosh(\pi\omega/2)\sinh(\pi\omega/h)}.\label{33}
\end{eqnarray}
The BA equations with the kernel (\ref{33}) were studied in \cite{7}, where the function ${\cal{E}}_\chi(A)$ corresponding to this kernel was calculated. It can be also written in the parametric form:
\begin{eqnarray}
&&{\cal{E}}_{\chi}(A)-{\cal{E}}(0)=-\frac{2A^2}{\beta^2}\Big[1-2u-hu(1-u)\big(1-\big(\frac{u}{1-u}\big)^{2/h}\big)\Big]+{O}(1)\nonumber \\ 
&&\ \ \ \ \ \ \ \  \Big(\frac{M}{2A}\Big)^2=u(1-u)\big(\frac{u}{1-u}\big)^{2/h}.\label{34}
\end{eqnarray}

It was shown in \cite{7} that function (\ref{34}) coincides with the classical minimum of the functional ${\cal{H}}_{\chi}$ (\ref{21}), where in the weak coupling limit we can neglect the fermionic terms. This gives us an additional test relating QFT (\ref{1}) and FST (\ref{7}). The singular behavior (${O}(1/\beta^2)$) of function ${\cal{E}}_\chi$  reflects the instability of weakly coupled bosons with respect to introduction of external field. The threshold behavior $(\Delta=(A-M)/M << 1)$ of functions (\ref{32}) and (\ref{34}) is also rather different. The first function has there singularity $\sim \Delta^{3/2}$ characteristic for fermionic particles. The second one possesses there behavior $\sim \Delta^2$ characteristic for the weak coupling limit of bosonic theory. It follows from eq. (\ref{31}) that properties of particles $\psi$ and $\chi$ change drasticaly under the flow from weak to strong coupling (fermions transform to bosons and vice versa). 

The scattering theory of charged particles $\psi$ and $\chi$ (\ref{7}) is invariant under the transformation $\beta \rightarrow 4\pi/\beta$ together with fermion-boson transformation $\psi \leftrightarrow \chi$. The amplitudes $S_{a\psi}$ and $S_{a\chi}$ (\ref{19}) are invariant under this transformation. The amplitudes $S_{ab}(\theta)$ of scattering of neutral particles $M_a(\theta)$ are also self-dual. The remarkable property of the amplitudes $S_{ab}(\theta)$ is that with function $x(\beta)$ defined by eq. (\ref{29}) they coincide exactly with scattering matrix of particles in pure $BC_n$ ATT proposed in ref. \cite{4}. The pure $BC_n$ ATT can be obtained from the QFT (\ref{1}) by the reduction $\psi=\chi=0$. It means that all contributions to the amplitudes $S_{ab}$ of neutral particles coming from charged particles $\chi$ and $\psi$ exactly cancel each other. This cancellation (which can be checked in perturbation theory) can not be occasionnal. It should be a symmetry responsible for this compensation property. We discuss this symmetry in the next section.
 
\section{Concluding remarks}
\ \ \ \ \ The symmetry responsible for the exact compensation of fermion and boson contributions to the amplitudes of neutral particles should relate $\psi$ and $\chi$ particles. It is possible only if this symmetry is generated by the conserved charges with half-integer spins. It should also be consistent with the FST (\ref{7}). One can check that this scattering matrix commutes with the symmetry algebra ${\mathfrak{T}}_n$, generated by the charges $Q_{\pm}(\overline{Q}_{\pm})$ with (Lorentz) spin $s$ equal to \ $h/2=n+1/2$ \ $(-n-1/2)$, and ``fermion number'' $F$. The charges $Q_{\pm}, \overline{Q}_{\pm}, F$ possess the following commutation relations :
\begin{eqnarray}
&&Q^2_{\pm}=\overline{Q}^2_{\pm}=\lbrace Q_{+},\overline{Q}_{-}\rbrace=\lbrace Q_{-},\overline{Q}_{+}\rbrace=0\ ;\nonumber \\
&&\big[F, Q_{\pm}\big]=\pm Q_{\pm}\ ;\ \ \ \ \ \big[F, \overline{Q}_{\pm}\big]=\mp\overline{Q}_{\pm};\nonumber \\
&&\lbrace Q_{+},Q_{-}\rbrace=P_h\ ; \ \ \ \ \ \lbrace \overline{Q}_{+},\overline{Q}_{-}\rbrace=\overline{P}_h,\label{35}
\end{eqnarray}
where $P_h(\overline{P}_h)$ is the right (left) component of the local integral of motion with spin $h$. This local charge acts on asymptotic (in, out) states of charged particles $\psi(\theta)$ and $\chi(\theta)$ with the eigenvalue $\lambda^2(\theta)$ $(\overline{\lambda}^2(\theta))$ :
\begin{eqnarray}
\lambda(\theta)=(M\exp\theta)^{h/2}\ ;\ \ \ \ \overline{\lambda}(\theta)=(M\exp\theta)^{-h/2}.\label{36}
\end{eqnarray}
The action of the operators $Q_{\pm}(\overline{Q}_{\pm})$ and $e^{i\pi F}$ on the one-particle states are defined as :
\begin{eqnarray}
&&Q_+|\psi(\theta)>=\lambda(\theta)|\overline{\chi}(\theta)>\ ; \ \ \ Q_+|\chi(\theta)>=\lambda(\theta)|\psi^{\dagger}(\theta)>\ ; \nonumber \\
&&Q_-|\psi^{\dagger}(\theta)>=\lambda(\theta)|\chi(\theta)>\ ; \ \ \ Q_-|\overline{\chi}(\theta)>=\lambda(\theta)|\psi(\theta)>\ ;\label{37} \\
&&e^{i\pi F}|\chi(\theta)>=e^{i\pi x}|\chi(\theta)>\ ; \ \ \ e^{i\pi F}|\psi(\theta)>=-e^{-i\pi x}|\psi(\theta)>\ ; \nonumber \\
&&e^{i\pi F}|\overline{\chi}(\theta)>=e^{-i\pi x}|\overline{\chi}(\theta)>\ ; \ \ \ e^{i\pi F}|\psi^{\dagger}(\theta)>=-e^{i\pi x}|\psi^{\dagger}(\theta)>. \nonumber 
\end{eqnarray}
The operators $\overline{Q}_{\mp}$ act at the same way as $Q_\pm$ with the substitution $\lambda \rightarrow \overline{\lambda}$. The action of the charges $Q_{\pm}(\overline{Q}_{\pm})$ on many particles sates can be defined from the following co-product rules :
\begin{eqnarray}
&&\Delta(Q_{\pm}) = Q_{\pm}\otimes I + e^{\pm i\pi F}\otimes Q_{\pm},\nonumber \\
&&\Delta(\overline{Q}_{\mp}) = {\overline Q}_{\mp}\otimes I + e^{\pm i\pi F}\otimes {\overline Q}_{\mp}.\label{38}
\end{eqnarray}
Using eqs. (\ref{37}), (\ref{38}) and commutativity condition of algebra ${\mathfrak{T}}_n$ with S-matrix (\ref{4}), we can derive all the ratios of the amplitudes (\ref{7}).

The eigenvalues of the operator $\exp(i\pi F)$ on the states $|\chi>(|\psi>)$ move from $1$ $(-1)$ at $\beta=0$, to $-1$ $(1)$ at $\beta=\infty$. Particles $\chi$\ $(\psi)$ have fractional values of fermion number equal to $x(\beta)$\ $(1-x(\beta))$. At the ``self-dual'' point $\beta^2=4\pi$ the fermion numbers of particles $\psi$ and $\chi$ are equal to $1/2$. At this point we have the symmetry $\psi \leftrightarrow \chi$. The amplitudes $S_\psi$ and $S_\chi$ coincide and S-matrix (\ref{7}) of the QFT ${\cal{BC}}_n(\psi,\chi,\sqrt{4\pi})$ can be expressed as the direct product of two-particle S-matrices of sine-Gordon model :
\begin{eqnarray}
S(\theta)=\prod_{a=0}^{n}\frac{\sinh\theta - i\sin(\pi a/h)}{\sinh\theta + i\sin(\pi a/h)}S_n(\theta)\otimes S_n(\theta)
\end{eqnarray}
where $S_n(\theta)$ is S-matrix of sine-Gordon model corresponding to the coupling constant $\beta^2_{SG}/8\pi=\frac{2}{2n+3}$. The FST (39) coincides with S-matrix of ${\cal{C}}_{n+1}^{(1)}(\psi_{1,2},\sqrt{4\pi})$ model considered in \cite{7}. It means that corresponding QFTs also coincide at the point $\beta^2=4\pi$.

To construct the currents $T_{n+3/2}^{(\pm)}(\overline{T}_{n+3/2}^{(\mp)})$ generating conserved charges $Q_{\pm}(\overline{Q}_{\mp})$ it is convenient to rewrite the action (\ref{1}) in terms of other fields. We introduce the scalar field $\Phi$ related with the fields $\psi, \psi^{\dagger}$ by the usual bosonisation rules \cite{1}, and the fields $\Phi_0, \varphi_0$ which describe the dual representation for CSG-model \cite{7,11}. In terms of these fields the action (\ref{1}) can be represented as :
\begin{eqnarray} 
{\cal A}_n &=&  \int d^2x\ \Big[\frac{1}{2}(\partial_{\mu}\Phi_0)^2 + \frac{1}{2}(\partial_{\mu}\varphi_0)^2 - M_0\cos(\alpha'\Phi_0) e^{\beta'\varphi_0} \label{40} \\
&&\ \ \ \ \ \ \ \ \ \ \ \ \ + \frac{1}{2}(\partial_{\mu}\Phi)^2 - M_0\cos(\alpha\Phi) e^{-\beta\varphi_n} + \frac{1}{2}(\partial_{\mu}\varphi)^2 - \frac{M_0^2}{2\beta^2}\big[ \sum_{i=0}^{n-1}\exp\beta(\varphi_{i+1}-\varphi_{i}) \big] \Big],\nonumber
\end{eqnarray}
where parameters $\alpha, \alpha', \beta' $ are defined by the relations : 
\begin{eqnarray} 
\alpha^2-\beta^2=4\pi\ , \ \ {\alpha'}^2-{\beta'}^2=4\pi\ , \ \ \beta'=4\pi/\beta.\label{41}
\end{eqnarray}
The first three terms in (\ref{40}) which we denote as $A_\sigma(\Phi_0,\varphi_0)$ correspond to the first (sigma-model) term in the action (\ref{1}). With action $A_\sigma(\Phi_0,\varphi_0)$ the fields $\chi, \overline{\chi}$ and $\chi\overline{\chi}$ can be represented as :
\begin{eqnarray}
&&\chi \sim \exp\big(\frac{2i\pi}{\alpha'}\Phi_0-\frac{2\pi}{\beta'}\varphi_0\big)\ ;\ \ \ 
\overline{\chi} \sim \exp\big(-\frac{2i\pi}{\alpha'}\Phi_0-\frac{2\pi}{\beta'}\varphi_0\big)\ ; \label{42} \\
&&\chi\overline{\chi} - const \sim \exp\big(-\frac{4\pi}{\beta'}\varphi_0\big)=\exp(-\beta\varphi_0).\nonumber
\end{eqnarray}
The action (\ref{40}) is not available for the ordinary perturbation theory in coupling constant $\beta$, however with this action QFT ${\cal{BC}}_n(\psi,\chi,\sqrt{4\pi})$ can be treated as perturbed conformal field theory. Using this approach we introduce the fields $\phi_0(\overline{\phi_0})$, $\phi(\overline{\phi})$ which are the right (left) chiral components of fields $\Phi_0, \Phi$, and right (left) derivatives $\partial=\partial_0+\partial_1\  (\overline{\partial}=\partial_0-\partial_1)$. The straightforward calculation shows that spin $n+3/2 \ (-(n+3/2))$ currents $T_{n+3/2}^{(\pm)}(\overline{T}_{n+3/2}^{(\mp)})$, which generate conserved charges $Q_{\pm}(\overline{Q}_{\mp})$can be written in the form :
\begin{eqnarray}
&&T_{n+3/2}^{(\pm)}=\exp\big(\pm \frac{4\pi i}{\alpha}\phi\big)(\kappa\partial-\partial\varphi_n)...(\kappa\partial-\partial\varphi_1)(\kappa\partial-\partial\varphi_0)\exp\big(\pm \frac{4\pi i}{\alpha'}\phi_0\big), \label{43}\\
&&\mbox{where} \ \ \ \kappa=\big(\frac{\beta}{4\pi}+\frac{1}{\beta}\big).\nonumber
\end{eqnarray}
The currents $\overline{T}_{n+3/2}^{(\mp)}$ can be obtained from the eq. (\ref{43}) by the substitution $\phi_0 \rightarrow \overline{\phi_0}$, $\phi \rightarrow \overline{\phi}$, $\partial \rightarrow \overline{\partial}$.

The fermion number $F$ can be expressed through the charges $Q_\psi$ and $Q_\chi$ (\ref{2}) as :
\begin{eqnarray}
F=x(\beta)Q_\chi + (1-x(\beta))Q_\psi\ .\label{44}
\end{eqnarray}
Together with charges $Q_\pm$ and $\overline{Q}_\mp$ it generates the algebra ${\mathfrak{T}}_n$. It follows from the representation (\ref{15}) for the neutral particles $M_a(\theta)$ and eq. (\ref{36}) that all these particles are annihilated by the local conserved charge $P_h$ :
\begin{eqnarray}
P_h|M_{a_1}(\theta_1)...M_{a_n}(\theta_n)>_{in(out)}=0.\label{45}
\end{eqnarray}
The neutral boson sector of the theory $\mathfrak{N}$, which contains only asymptotic particles $M_a(\theta)$, is defined by the kernel of the conserved charge $P_h$. The restriction of the scattering theory to the neutral sector $\mathfrak{N}$, defines self-consistent FST. The QFT corresponding to this FST possesses the Lagrangian description with the local action, which can be obtained from the action (\ref{1}) by the reduction $\psi=\chi=0$.

At the end we note that besides this reduction, the QFT (\ref{1}) possesses several other integrable reductions. For example : $\chi=0$, $\psi \neq 0$; $\psi=0$, $\chi \neq 0$; $\psi \neq 0$, $\chi$ is real; $\chi \neq 0$, $\psi$ is Majorana fermion; $\chi$ is real, $\psi$ is Majorana fermion and so on. The FSTs corresponding to these reductions are described in refs. \cite{4,7,8}.
\\
\\


\begin{thebibliography}{10}
\bibitem{1}
S. Coleman, Phys. Rev. D 11 (1975) 2088;\\
S. Mandelstam, Phys. Rev. D 11 (1975) 3026.
%
\bibitem{2}
C. Montonen and D. Olive, Phys. Lett. B 78 (1977) 117;\\
P. Goddard, J. Nuyts and D. Olive, Nucl. Phys. B 125 (1977) 1.
%
\bibitem{3}
N. Seiberg and E. Witten, Nucl. Phys. B 426 (1994) 19; Nucl. Phys. B 431 (1994) 484.
%
\bibitem{4}
H.W. Braden, E. Corrigan, P.E. Dorey and R. Sasaki, Nucl. Phys. B 338 (1990) 689;\\
G.W. Delius, M.T. Grisaru and D. Zanon, Nucl. Phys. B 382 (1992) 365.
%
\bibitem{5}
K. Pohlmayer, Comm. Math. Phys. 46 (1976) 207:\\
F. Lund and T. Regge, Phys. Rev. D 14 (1976) 1524;\\
H. J. de Vega and J. M. Maillet, Phys. Lett. B 101 (1981) 302; Phys. Rev. D 28 (1983) 1441;\\
C. Bonneau and F. Delduc, Nucl. Phys. B 245 (1985) 561. 
%
\bibitem{6}
E. Witten, Phys. Rev. D 44 (1991) 314.
%
\bibitem{7}
V.A. Fateev, Nucl.Phys. B 479 (1996) 594.
%
\bibitem{8}
G.W. Delius, M.T. Grisaru and D. Zanon,Phys. Lett. B 256 (1991) 164;\\
C. Destri, H.J. De Vega and V.A. Fateev, Phys. Lett. B 256 (1991) 173.
%
\bibitem{9}
G. Japaridze, A. Nersesyan and P. Wieghmann, Nucl. Phys. B 230 (1984) 511;\\
P. Hasenfratz, M. Maggiore and F. Niedermayer, Phys. Lett. B 245 (1990) 522;\\
Al. Zamolodchikov, Int. J. Mod. Phys. A 10 (1995) 1125.
%
\bibitem{10}
V.A. Fateev, E. Onofri and Al. Zamolodchikov, Nucl. Phys. B 406 (1993) 521.
%
\bibitem{11}
V.A. Fateev, Phys. Lett. B 357 (1995) 397.

\end{thebibliography}
\end{document}